\documentclass[aps,preprint,showpacs,groupedaddress]{revtex4-1}
\usepackage{amssymb}
\usepackage{mathrsfs}
\usepackage{amsmath}
\usepackage{mathtools}
\usepackage{pstricks}
\usepackage{color}
\usepackage{graphicx}
\usepackage{amsthm}

\usepackage{physics}
\begin{document}

%\preprint{}

\title{\bf Critical properties of scalar field theory with Lorentz violation: Exact treatment of Lorentz-violating mechanism}


%\author{William C. Vieira}
%\email{william.vieira@gmail.com}
%\affiliation{\it Departamento de F\'\i sica, Universidade Federal do Piau\'\i, 64049-550, Teresina, PI, Brazil}

\author{P. R. S. Carvalho}
\email{prscarvalho@ufpi.edu.br}
\affiliation{\it Departamento de F\'\i sica, Universidade Federal do Piau\'\i, 64049-550, Teresina, PI, Brazil}

\author{M. I. Sena-Junior}
\email{marconesena@poli.br}
\affiliation{\it Escola Polit\'{e}cnica de Pernambuco, Universidade de Pernambuco, 50720-001, Recife, PE, Brazil}
\affiliation{\it Instituto de F\'{i}sica, Universidade Federal de Alagoas, 57072-900, Macei\'{o}, AL, Brazil}

%\thanks{}


%\date{\today}

\begin{abstract}
In this work, we compute analytically the infrared divergences of massless O($N$) self-interacting scalar field theories with Lorentz violation, which are exact in the Lorentz-violating $K_{\mu\nu}$ coefficients, for evaluating the corresponding next-to-leading order critical exponents. For that, we apply three distinct and independent field-theoretic renormalization group methods. We find that the outcomes for the critical exponents are the same in the three methods and, furthermore, are identical to their Lorentz invariant counterparts. We generalize the results for all loop levels by employing a general theorem arising from the exact procedure and give the corresponding physical interpretation. 
\end{abstract}

%\pacs{11.30.-j; 64.60.ae; 11.30.Cp}
%\keywords{}

\maketitle


\section{Introduction} 

\par The search for violations of Lorentz symmetry has grown in the last few years. The possible physical effects predicted by Lorentz-violating (LV) theories were established in research areas ranging from high energy physics
\cite{Universe2201630,PhysRevD40198918863,PhysRevD391989683,PhysRevD3591991545,PhysRevD59199124021,
LivingRevRelativ1620135,AmelinoCamelia,PhysRevLett902013211601,PhysLettB478200039,PhysRevLett872001141601,
PhysRevD672003043508,PhysRevD92045016,ApJ806269,IntJModPhysA3015500729,PhysLettB742236,PhysRevD86125015,
PhysRevD.84.065030,Carvalho2014320,Carvalho2013850} to condensed matter theory \cite{EurophysLett.108.21001,Int.J.Mod.Phys.B.30.1550259,Int.J.Geom.MethodsMod.Phys.13.1650049}. Except to one work considering the breaking of the referred symmetry exactly \cite{PhysRevLett832518}, for our knowledge, the most of them treat this symmetry breaking mechanism in an approximated footing at first order \cite{PhysRevD92045016,ApJ806269,IntJModPhysA3015500729,PhysLettB742236,PhysRevD86125015,Carvalho2014320,PhysRevD58116002,
PhysRevD65056006,PhysRevD79125019} and just a few of them at second order \cite{PhysRevD.84.065030,Carvalho2013850,EurophysLett.108.21001,Int.J.Mod.Phys.B.30.1550259,Int.J.Geom.MethodsMod.Phys.13.1650049} in a typical LV parameter. All these works were treated perturbatively in two parameters, some LV one and the loop level considered. While these problems were investigated at most at second order in the LV parameter, the same was not the case when dealing with the loop level in question. In fact, after an explicit analytic computation up to two- and three-loop order for the critical exponents $\nu$ and $\eta$, respectively, some degree of sophistication in a road to an all-loop order solution of the problem was attained when expressions for the same exponents at any loop level were obtained, although yet at second order in the LV parameter for $\nu$ and at first order in the same parameter for $\eta$. We present an exact solution of the problem valid for any values of LV parameter.

\par  As the critical exponents are universal quantities, they do not depend on the microscopic details of the system but depend only on its dimension $d$, $N$ and symmetry of some $N$-component order parameter (magnetization for magnetic systems) if the interactions of its constituents are of short- or long-range type. In this work, we propose to probe the effect of exact Lorentz symmetry breaking mechanism in the outcomes for the all-loop critical exponents for massless O($N$) self-interacting scalar field theories with Lorentz violation. For that, we apply three distinct field-theoretic methods based on renormalization group and $\epsilon$-expansion techniques. The systems studied here belong to the general O($N$) universality class and are distinct systems as a fluid and a ferromagnet, whose critical behaviors are characterized by the same set of critical exponents. The referred universality class is a generalization of the specific models with short-range interactions: Ising ($N=1$), XY ($N=2$), Heisenberg ($N=3$), self-avoiding random walk ($N=0$), spherical ($N \rightarrow \infty$) etc \cite{Pelissetto2002549}. In the field-theoretic formulation, the universality hypothesis implies that the final values for the critical exponents must be the same if they are computed in distinct theories renormalized at different renormalization schemes, although these theories are different at intermediate steps and consequently display distinct renormalization constants, $\beta$-functions, anomalous dimensions, fixed points etc. The critical exponents are obtained through the scaling properties of the primitively $1$PI vertex parts of the theory, namely the $\Gamma^{(2)}$, $\Gamma^{(4)}$ and $\Gamma^{(2,1)}$ ones. These vertex parts are associated to the correlation functions of the system and contain all divergent properties of the theory. From the parameters whose critical exponents depend, the symmetry of order parameter is the less intuitive one. Investigating the influence of this parameter on the values of the critical exponents is one of the aims of this work. Another one is to illustrate how the symmetry breaking mechanism approached here is absorbed into a Lorentz-invariant effective theory. In fact, this Lorentz violation is only a naively one. This fact is proved through coordinates redefinition techniques such that the metric of the LV theory is converted into the metric of the usual Lorentz invariant theory. These coordinates redefinition techniques were presented originally by Kostelecky \cite{PhysRevD692004105009}. A detailed exposition is given by Kostelecky and Tasson \cite{PhysRevD832011016013}.

\section{Exact Lorentz-violating next-to-leading order critical exponents in the normalization conditions method}

\par In the normalization condition method, we start from the bare theory. The one studied here is represented by the bare Lagrangian density 
\begin{eqnarray}\label{huytrji}
\mathcal{L}_{B} = \frac{1}{2}\partial^{\mu}\phi_{B}\partial_{\mu}\phi_{B} + \frac{1}{2}K_{\mu\nu}\partial^{\mu}\phi_{B}\partial^{\nu}\phi_{B} + \frac{\lambda_{B}}{4!}\phi_{B}^{4} + \frac{1}{2}t_{B}\phi_{B}^{2},
\end{eqnarray}
where the Lorentz symmetry is violated if the LV coefficients are chosen such that they do not transform as a second order tensor under Lorentz transformations, i.e. $K_{\mu\nu}^{\prime} \neq \Lambda_{\mu}^{\rho}\Lambda_{\nu}^{\sigma} K_{\rho\sigma}$ and is renormalized through the multiplicative renormalization of the primitively $1$PI vertex parts $\Gamma^{(n, l)}(P_{i}, Q_{j}, g, \kappa) = Z_{\phi}^{n/2}Z_{\phi^{2}}^{l}\Gamma_{B}^{(n, l)}(P_{i}, Q_{j}, \lambda_{B})$ ($i = 1, \cdots, n$, $j = 1, \cdots, l$) by fixing their external momenta at convenient values through the normalization conditions
\begin{eqnarray}\label{ygfdxzsze}
\Gamma^{(2)}(P^{2} + K_{\mu\nu}P^{\mu}P^{\nu} = 0, g) = 0, 
\end{eqnarray}
\begin{eqnarray}
\frac{\partial \Gamma^{(2)}(P^{2} + K_{\mu\nu}P^{\mu}P^{\nu}, g)}{\partial (P^{2} + K_{\mu\nu}P^{\mu}P^{\nu})}\Biggr|_{P^{2} + K_{\mu\nu}P^{\mu}P^{\nu} = \kappa^{2}}   = 1,
\end{eqnarray}
\begin{eqnarray}\label{jijhygtfrd}
\Gamma^{(4)}(P^{2} + K_{\mu\nu}P^{\mu}P^{\nu}, g)|_{SP} = g, 
\end{eqnarray}
\begin{eqnarray}
\Gamma^{(2,1)}(P_{1}, P_{2}, Q_{3}, g)|_{\overline{SP}} = 1,
\end{eqnarray}
where  $Q_{3} = -(P_{1} + P_{2})$ and for SP: $P_{i}\cdot P_{j} = (\kappa^{2}/4)(4\delta_{ij}-1)$, implying that $(P_{i} + P_{j})^{2} \equiv P^{2} + K_{\mu\nu}P^{\mu}P^{\nu} = \kappa^{2}$ for $i\neq j$ and for $\overline{SP}$: $P_{i}^{2} = 3\kappa^{2}/4$ and $P_{1}\cdot P_{2} = -\kappa^{2}/4$, implying $(P_{1} + P_{2})^{2} \equiv P^{2} + K_{\mu\nu}P^{\mu}P^{\nu} = \kappa^{2}$ and $\kappa$ is an arbitrary momentum scale parameter. We are defining the dimensionless bare $u_{0}$ and renormalized $u$ coupling constants as $\lambda_{B} = u_{B}\kappa^{\epsilon/2}$ and $g = u\kappa^{\epsilon/2}$, respectively, where $\epsilon = 4 - d$. The LV coefficients $K_{\mu\nu}$ are dimensionless, symmetric ($K_{\mu\nu} = K_{\nu\mu}$) and are the same for all $N$ components of the field. Furthermore they preserve the O($N$) symmetry of the $N$-component field. The quantities $\phi_{B}$, $\lambda_{B}$ and $t_{B}$ are the bare field, coupling constant and composite field coupling constant, respectively. For next-to-leading order, the primitively $1$PI vertex parts are given by
\begin{eqnarray}\label{gtfrdrdes}
\Gamma^{(2)}_{B} = \quad \parbox{12mm}{\includegraphics[scale=1.0]{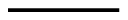}}^{-1} + \frac{1}{6}\hspace{1mm}\parbox{12mm}{\includegraphics[scale=1.0]{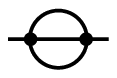}} + \frac{1}{4}\hspace{1mm}\parbox{10mm}{\includegraphics[scale=0.8]{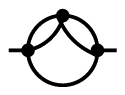}} ,
\end{eqnarray}
\begin{eqnarray}
\Gamma^{(4)}_{B} = \quad \parbox{12mm}{\includegraphics[scale=0.09]{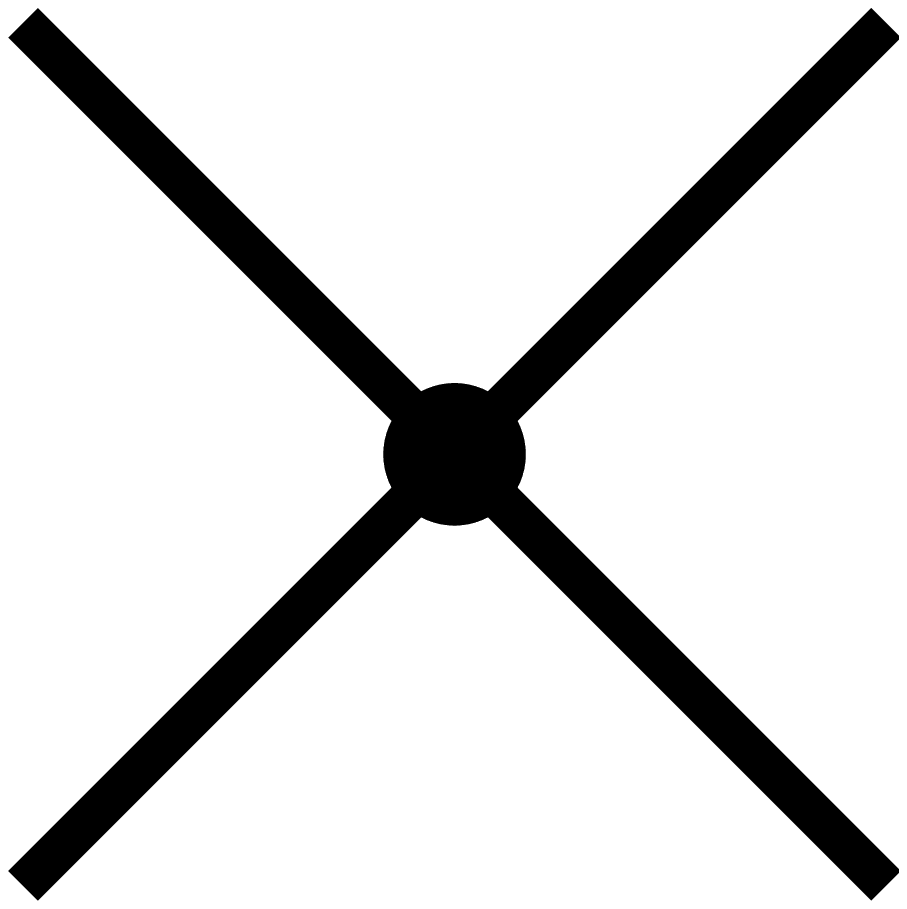}} + \quad \frac{1}{2}\hspace{1mm}\parbox{10mm}{\includegraphics[scale=1.0]{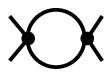}} + 2 \hspace{1mm} perm. \quad + \frac{1}{4}\hspace{1mm}\parbox{16mm}{\includegraphics[scale=1.0]{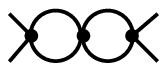}} + 2 \hspace{1mm} perm. \quad +  \frac{1}{2}\hspace{1mm}\parbox{12mm}{\includegraphics[scale=0.8]{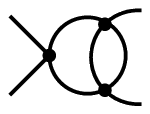}} + 5 \hspace{1mm} perm. , 
\end{eqnarray}
\begin{eqnarray}\label{gtfrdesuuji}
\Gamma^{(2,1)}_{B} = \quad \parbox{14mm}{\includegraphics[scale=1.0]{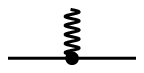}} \quad + \quad \frac{1}{2}\hspace{1mm}\parbox{14mm}{\includegraphics[scale=1.0]{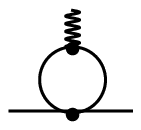}} \quad + \frac{1}{4}\hspace{1mm}\parbox{12mm}{\includegraphics[scale=1.0]{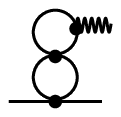}} \quad + \quad \frac{1}{2}\hspace{1mm}\parbox{12mm}{\includegraphics[scale=0.8]{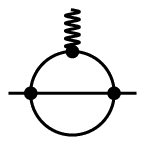}} ,
\end{eqnarray}
where \textit{perm.} means a permutation of the external momenta attached to the cut external lines. Each internal line represents the free LV Green's function $G_{0}^{-1}(q) = \parbox{12mm}{\includegraphics[scale=1.0]{fig9.eps}}^{-1} = q^{2} + K_{\mu\nu}q^{\mu}q^{\nu}$. Now we can express the renormalization constants $Z_{\phi}$ and $\overline{Z}_{\phi^{2}} \equiv Z_{\phi}Z_{\phi^{2}}$ in terms of powers of $u$ perturbatively and compute the $\beta$-function and anomalous dimensions present in the renornalization group equation of the theory
\begin{eqnarray}
\left( \kappa\frac{\partial}{\partial\kappa} + \beta\frac{\partial}{\partial u} - \frac{1}{2}n\gamma_{\phi} + l\gamma_{\phi^{2}} \right)\Gamma_{R}^{(n, l)} = 0\quad\quad
\end{eqnarray}
where 
\begin{eqnarray}\label{kjsjffbxdzs}
\beta(u) = \kappa\frac{\partial u}{\partial \kappa} = -\epsilon\left(\frac{\partial\ln u_{0}}{\partial u}\right)^{-1},
\end{eqnarray}
\begin{eqnarray}\label{koiuahygtf}
\gamma_{\phi}(u) = \beta(u)\frac{\partial\ln Z_{\phi}}{\partial u},
\end{eqnarray}
\begin{eqnarray}\label{koiubhygtf}
\gamma_{\phi^{2}}(u) = -\beta(u)\frac{\partial\ln Z_{\phi^{2}}}{\partial u}.
\end{eqnarray}
We will use the function
\begin{eqnarray}\label{udgygeykoiuchygtf}
\overline{\gamma}_{\phi^{2}}(u) = -\beta(u)\frac{\partial\ln \overline{Z}_{\phi^{2}}}{\partial u} \equiv \gamma_{\phi^{2}}(u) - \gamma_{\phi}(u)
\end{eqnarray}
for convenience. The exact evaluation in the LV coefficients proceeds by 
%noting that $q^{2} + K_{\mu\nu}q^{\mu}q^{\nu} \equiv (\delta_{\mu\nu} + K_{\mu\nu})q^{\mu}q^{\nu}$ = $q^{t}(\mathbb{I}$ + $\mathbb{K})q$, where $q$ is a $d$-dimensional vector whose representation is a column matrix and $q^{t}$ is a row one and $\mathbb{I}$ and $\mathbb{K}$ are matrix representations of the identity and $K_{\mu\nu}$, respectively. Thus through the 
changing the variables \cite{CarvalhoSenaJunior} through coordinates redefinition in momentum space directly in Feynman diagrams where $q^{\prime} = \sqrt{\mathbb{I} + \mathbb{K}}\hspace{1mm}q$.
As it is known, from all diagrams displayed above, we need to compute only four of them \cite{Amit}, namely, for $N = 1$,
\begin{eqnarray}
\parbox{10mm}{\includegraphics[scale=1.0]{fig10.eps}}_{SP} = \frac{1}{\epsilon}\left(1 + \frac{1}{2}\epsilon  \right)\mathbf{\Pi}, 
\end{eqnarray}   
\begin{eqnarray}
\parbox{12mm}{\includegraphics[scale=1.0]{fig6.eps}}^{\prime} = -\frac{1}{8\epsilon}\left( 1 + \frac{5}{4}\epsilon \right)\mathbf{\Pi}^{2},
\end{eqnarray}  
\begin{eqnarray}
\parbox{10mm}{\includegraphics[scale=0.9]{fig7.eps}}^{\prime} = -\frac{1}{6\epsilon^{2}}\left( 1 + 2\epsilon  \right)\mathbf{\Pi}^{3},
\end{eqnarray}  
\begin{eqnarray}
\parbox{12mm}{\includegraphics[scale=0.8]{fig21.eps}}_{SP} = \frac{1}{2\epsilon^{2}}\left( 1 + \frac{3}{2}\epsilon \right)\mathbf{\Pi}^{2},
\end{eqnarray}  
where, $\mathbf{\Pi} = 1/\sqrt{det(\mathbb{I} + \mathbb{K})}$ is a LV full factor, $\parbox{6mm}{\includegraphics[scale=0.6]{fig10.eps}}_{SP} \equiv \parbox{6mm}{\includegraphics[scale=0.6]{fig10.eps}}\vert_{P^{2} + K_{\mu\nu}P^{\mu}P^{\nu} = 1}$, $\parbox{8mm}{\includegraphics[scale=0.7]{fig6.eps}}^{\prime} \equiv [\partial \parbox{8mm}{\includegraphics[scale=0.7]{fig6.eps}}/\partial ( P^{2} + K_{\mu\nu}P^{\mu}P^{\nu})]\vert_{P^{2} + K_{\mu\nu}P^{\mu}P^{\nu} = 1}$, $\parbox{8mm}{\includegraphics[scale=0.5]{fig21.eps}}_{SP} \equiv \parbox{8mm}{\includegraphics[scale=0.5]{fig21.eps}}\vert_{P^{2} + K_{\mu\nu}P^{\mu}P^{\nu} = 1}$ and $\parbox{8mm}{\includegraphics[scale=0.6]{fig7.eps}}^{\prime} \equiv [\partial \parbox{8mm}{\includegraphics[scale=0.6]{fig7.eps}}/\partial (P^{2} + K_{\mu\nu}P^{\mu}P^{\nu})]\vert_{P^{2} + K_{\mu\nu}P^{\mu}P^{\nu} = 1}$ and we have written all momenta in the diagrams in units of $\kappa$ and absorbed $\kappa$ in the dimensionless renormalized coupling constant, thus turning out the momenta as dimensionless quantities.  The symmetry point now is given by $P^{2} + K_{\mu\nu}P^{\mu}P^{\nu} = \kappa^{2} \rightarrow 1$. Now computing the $\beta$-function and anomalous dimensions for general $N$, we obtain 
\begin{eqnarray}\label{fhufhudh}
\beta(u) = -\epsilon u +   \frac{N + 8}{6}\left( 1 + \frac{1}{2}\epsilon \right)\mathbf{\Pi} u^{2} -  \frac{3N + 14}{12}\mathbf{\Pi}^{2}u^{3}, 
\end{eqnarray}
\begin{eqnarray}\label{gkjlhitu}
\gamma_{\phi}(u) = \frac{N + 2}{72}\left( 1 + \frac{5}{4}\epsilon \right)\mathbf{\Pi}^{2}u^{2} - \frac{(N + 2)(N + 8)}{864}\mathbf{\Pi}^{3}u^{3},  
\end{eqnarray}
\begin{eqnarray}\label{dkvyenh}
\overline{\gamma}_{\phi^{2}}(u) = \frac{N + 2}{6}\left( 1 + \frac{1}{2}\epsilon \right)\mathbf{\Pi} u -  \frac{N + 2}{12}\mathbf{\Pi}^{2}u^{2}.
\end{eqnarray}
The present approach, the coordinates redefinition in momentum space directly in Feynman diagrams one, besides to be exact, gives correct expressions for the $\beta$-function and anomalous dimensions in an evident and direct way based on a single concept, that of loop order of the referred term of the corresponding function. For example, as the first term of the $\beta$-function of Eq. (\ref{fhufhudh}) is not associated to any loop integral, according to the present approach, it must not have to acquire a LV full $\mathbf{\Pi}$ factor, although it is of first order in $u$. The second one, although being of second order in $u$, must acquire one power of the LV full $\mathbf{\Pi}$ factor. The third term is of third order in $u$ and must be of second order in $\mathbf{\Pi}$, since it is a two-loop order one. The same argument can be applied to the others terms of the anomalous dimensions of Eqs. (\ref{gkjlhitu})-(\ref{dkvyenh}) as well. We also observe that the $\beta$-function and anomalous dimensions, in this method, depend on the LV coefficients at its exact form only through the LV $\mathbf{\Pi}$ factor. Now, for the evaluation of the radiative quantum corrections or equivalently the corrections to mean field or Landau approximation to the critical exponents, we need to compute the nontrivial solution of the $\beta$-function (the trivial one leads to the mean field or Landau critical exponents) 
\begin{eqnarray}
u^{*} = \frac{6\epsilon}{(N + 8)\mathbf{\Pi}} \left\{ 1 + \epsilon\left[ \frac{3(3N + 14)}{(N + 8)^{2}} -\frac{1}{2} \right]\right\}.
\end{eqnarray}
We can see that as the first term of the $\beta$-function does not contain a LV full $\mathbf{\Pi}$ factor, by factoring $u$ from that function, we have that in this method the nontrivial fixed point is given by $u^{*} = u^{*(0)}/\mathbf{\Pi}$, where $u^{*(0)}$ is its Lorentz-invariant (LI) counterpart. Now by applying the relations $\eta\equiv\gamma_{\phi}(u^{*})$ and $\nu^{-1}\equiv 2 - \eta - \overline{\gamma}_{\phi^{2}}(u^{*})$
we find that the LV critical exponents are identical to that of the corresponding LI theory, namely $\eta\equiv\eta^{(0)}$ and $\nu\equiv\nu^{(0)}$ \cite{Wilson197475,PhysRevLett.28.240,PhysRevLett.28.548}
\begin{eqnarray}\label{eta}
\eta = \frac{(N + 2)\epsilon^{2}}{2(N + 8)^{2}}\left\{ 1 + \epsilon\left[ \frac{6(3N + 14)}{(N + 8)^{2}} -\frac{1}{4} \right]\right\},
\end{eqnarray}
\begin{eqnarray}\label{nu}
\nu = \frac{1}{2} + \frac{(N + 2)\epsilon}{4(N + 8)} +  \frac{(N + 2)(N^{2} + 23N + 60)\epsilon^{2}}{8(N + 8)^{3}}.
\end{eqnarray}
Physically, the LV critical exponents are the same as their LI counterparts because the symmetry breaking mechanism occurs in the space where the field is defined and not in the internal space of the field, a fact which could change the values for the critical exponents. Furthermore, the field theory we are considering is only naively LV, as seen through coordinates redefinition techniques \cite{PhysRevD692004105009,PhysRevD832011016013}. Thus, the symmetry breaking mechanism approached here is absorbed into a Lorentz-invariant effective theory. As there are six critical exponents and four relations among them, then the two ones computed above are enough to evaluate the remaining ones. Now we proceed to compute the referred critical exponents in the minimal subtraction scheme.

\section{Exact Lorentz-violating next-to-leading order critical exponents in the minimal subtraction scheme}

\par In the minimal subtraction scheme, again we start from the bare theory. It is characterized by its generality and elegance, since, as opposed to the earlier renormalization scheme, the external momenta can assume any of their arbitrary values. As we do not fix the external momenta in the diagrams, the diagrams to be computed are the ones
\begin{eqnarray}
\parbox{10mm}{\includegraphics[scale=1.0]{fig10.eps}} = \frac{1}{\epsilon} \left[1 - \frac{1}{2}\epsilon - \frac{1}{2}\epsilon L(P^{2} + K_{\mu\nu}P^{\mu}P^{\nu}) \right]\mathbf{\Pi} ,
\end{eqnarray}   
\begin{eqnarray}
\parbox{12mm}{\includegraphics[scale=1.0]{fig6.eps}} = -\frac{P^{2} + K_{\mu\nu}P^{\mu}P^{\nu}}{8\epsilon}\left[ 1 + \frac{1}{4}\epsilon -2\epsilon L_{3}(P^{2} + K_{\mu\nu}P^{\mu}P^{\nu}) \right] \mathbf{\Pi}^{2},
\end{eqnarray}  
\begin{eqnarray}
\parbox{10mm}{\includegraphics[scale=0.9]{fig7.eps}} = -\frac{P^{2} + K_{\mu\nu}P^{\mu}P^{\nu}}{6\epsilon^{2}}\left[ 1 + \frac{1}{2}\epsilon -3\epsilon L_{3}(P^{2} + K_{\mu\nu}P^{\mu}P^{\nu}) \right] \mathbf{\Pi}^{3},
\end{eqnarray}  
\begin{eqnarray}
\parbox{12mm}{\includegraphics[scale=0.8]{fig21.eps}} = \frac{1}{2\epsilon^{2}}\left[1 - \frac{1}{2}\epsilon - \epsilon L(P^{2} + K_{\mu\nu}P^{\mu}P^{\nu}) \right]\mathbf{\Pi}^{2},
\end{eqnarray}
where
\begin{eqnarray}\label{uhduhguh}
L(P^{2} + K_{\mu\nu}P^{\mu}P^{\nu}) = \int_{0}^{1}dx\ln[x(1-x)(P^{2} + K_{\mu\nu}P^{\mu}P^{\nu})],
\end{eqnarray}
\begin{eqnarray}\label{uhduhguhf}
L_{3}(P^{2} + K_{\mu\nu}P^{\mu}P^{\nu}) = \int_{0}^{1}dx(1-x)\ln[x(1-x)(P^{2} + K_{\mu\nu}P^{\mu}P^{\nu})],
\end{eqnarray}
Now we can evaluate the $\beta$-function and anomalous dimensions and obtain
\begin{eqnarray}\label{uahuahuahu}
\beta(u) = -\epsilon u + \frac{N + 8}{6}\mathbf{\Pi}u^{2} - \frac{3N + 14}{12}\mathbf{\Pi}^{2}u^{3},
\end{eqnarray}
\begin{eqnarray}
\gamma_{\phi}(u) = \frac{N + 2}{72}\mathbf{\Pi}^{2}u^{2} - \frac{(N + 2)(N + 8)}{1728}\mathbf{\Pi}^{3}u^{3},
\end{eqnarray}
\begin{eqnarray}\label{uahuahuahuaa}
\overline{\gamma}_{\phi^{2}}(u) = \frac{N + 2}{6}\mathbf{\Pi} u - \frac{N + 2}{12}\mathbf{\Pi}^{2}u^{2}.
\end{eqnarray}
The present method shows its elegance when we see that the final form of the $\beta$-function and anomalous dimensions do not depend of the LV coefficients through the integrals (\ref{uhduhguh})-(\ref{uhduhguhf}), since they cancel out in the renormalization program and thus do not need to be evaluated. Thus the LV dependence of these function is only due to the LV full factor $\mathbf{\Pi}$. The nontrivial fixed point is given by 
\begin{eqnarray}
 u^{*} = \frac{6\epsilon}{(N + 8)\mathbf{\Pi}}\left\{ 1 + \epsilon\left[ \frac{3(3N + 14)}{(N + 8)^{2}} \right]\right\}.
\end{eqnarray}
And finally, evaluating the critical exponents, we obtain the same ones as their corresponding LI counterparts and arrive at the same results as that in the earlier method. This result confirms the universality hypothesis, whose field theoretic version asserts that universal physical quantities must be the same when computed in field theories renormalized at distinct renormalization schemes. This shows the great value of computing the critical exponents through distinct renormalization schemes. Now we have to evaluate the critical exponents by the application of a third method.

\section{Exact Lorentz-violating next-to-leading order critical exponents in the BPHZ method}

\par As opposed to the earliers methods, in the BPHZ (Bogoliubov-Parasyuk-Hepp-Zimmermann) method \cite{BogoliubovParasyuk,Hepp,Zimmermann}, we start from the renormalized theory at a given loop order. For example, if the theory is a bare one at its one-loop level, we eliminate the divergences by introducing some terms to the initial Lagrangian density such that a finite one is found. Then for the bare next loop level, we repeat the procedure and so on order by order in perturbation theory. The present method, for treating massless theories  \cite{Int.J.Mod.Phys.B.30.1550259,Amit}, can be easily confused with their massive counterpart \cite{CarvalhoSenaJunior}. There are many differences between them and we can present a few ones. The first of them is that in Eq. (\ref{Zphi}), the diagrams $\parbox{9mm}{\includegraphics[scale=.8]{fig6.eps}}$ and $\parbox{8mm}{\includegraphics[scale=.7]{fig7.eps}}$ depend on the mass in the latter method, while the same diagrams do not depend on $m$ in the former one. In Eq. (\ref{Zg}), besides the diagrams in question depend on the mass in the massive situation, there are two more diagrams, namely the $\parbox{8mm}{\includegraphics[scale=.8]{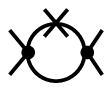}}$ and $\parbox{8mm}{\includegraphics[scale=.7]{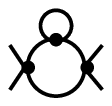}}$ ones. When we look at Eq. (\ref{Zphi2}), we can see the greater difference between the two methods, where in the massive theory, the $Z_{\phi^{2}}$ counterpart of $Z_{m^{2}}$ (which contains the following diagrams $\parbox{8mm}{\includegraphics[scale=.7]{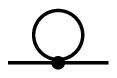}}$, $\parbox{8mm}{\includegraphics[scale=.7]{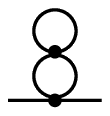}}$, $\parbox{8mm}{\includegraphics[scale=.7]{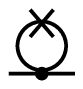}}$ and $\parbox{8mm}{\includegraphics[scale=.7]{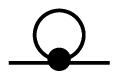}}$ for example) is composed of completely distinct diagrams of that of Eq.(\ref{Zphi2}). The divergences are absorbed by renormalization constants as
\begin{eqnarray}\label{huytrjii}
\mathcal{L} = \frac{1}{2}Z_{\phi}\partial^{\mu}\phi\partial_{\mu}\phi + \frac{1}{2}K_{\mu\nu}Z_{\phi}\partial^{\mu}\phi\partial^{\nu}\phi + \frac{\mu^{\epsilon}u}{4!}Z_{u}\phi^{4} + \frac{1}{2}tZ_{\phi^{2}}\phi^{2},
\end{eqnarray}
where
\begin{eqnarray}\label{huytr}
\phi = Z_{\phi}^{-1/2}\phi_{B},  
\end{eqnarray} 
\begin{eqnarray}\label{huytrd}
\mu^{-\epsilon}\frac{Z_{\phi}^{2}}{Z_{u}}\lambda_{B},  
\end{eqnarray} 
\begin{eqnarray}\label{huytrss}
t = \frac{Z_{\phi}}{Z_{\phi^{2}}}t_{B}. 
\end{eqnarray} 
The renormalization constants can be expanded as
\begin{eqnarray}\label{uhguhfgugu}
Z_{\phi} = 1 + \sum_{i=1}^{\infty} c_{\phi}^{i}, 
\end{eqnarray} 
\begin{eqnarray}\label{uhguhfgugud}
Z_{u} = 1 + \sum_{i=1}^{\infty} c_{u}^{i},
\end{eqnarray} 
\begin{eqnarray}\label{uhguhfgugus}
Z_{\phi^{2}} = 1 + \sum_{i=1}^{\infty} c_{\phi^{2}}^{i}
\end{eqnarray} 
The $c_{\phi}^{i}$, $c_{g}^{i}$ and $c_{\phi^{2}}^{i}$ coefficients are the $i$-th loop order renormalization constants for the field, renormalized coupling constant and composite field, respectively. They are given by
\begin{eqnarray}\label{Zphi}
&& Z_{\phi}(u,\epsilon^{-1}) = 1 +  \frac{1}{P^{2} + K_{\mu\nu}P^{\mu}P^{\nu}} \Biggl[ \frac{1}{6} \mathcal{K} 
\left(\parbox{12mm}{\includegraphics[scale=1.0]{fig6.eps}}
\right) S_{\parbox{10mm}{\includegraphics[scale=0.5]{fig6.eps}}} + \frac{1}{4} \mathcal{K} 
\left(\parbox{12mm}{\includegraphics[scale=1.0]{fig7.eps}} \right) S_{\parbox{6mm}{\includegraphics[scale=0.5]{fig7.eps}}} + \nonumber \\ && \frac{1}{3} \mathcal{K}
  \left(\parbox{12mm}{\includegraphics[scale=1.0]{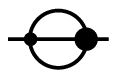}} \right) S_{\parbox{6mm}{\includegraphics[scale=0.5]{fig26.eps}}} \Biggr], \hspace{4mm}
\end{eqnarray}

\begin{eqnarray}\label{Zg}
&&Z_{u}(u,\epsilon^{-1}) = 1 + \frac{1}{\mu^{\epsilon}u} \Biggl[ \frac{1}{2} \mathcal{K} 
\left(\parbox{10mm}{\includegraphics[scale=1.0]{fig10.eps}} + 2 \hspace{1mm} perm.
\right) S_{\parbox{10mm}{\includegraphics[scale=0.5]{fig10.eps}}} + \frac{1}{4} \mathcal{K} 
\left(\parbox{17mm}{\includegraphics[scale=1.0]{fig11.eps}} + 2 \hspace{1mm} perm. \right) S_{\parbox{10mm}{\includegraphics[scale=0.5]{fig11.eps}}} \nonumber \\ && + \frac{1}{2} \mathcal{K} 
\left(\parbox{12mm}{\includegraphics[scale=.8]{fig21.eps}} + 5 \hspace{1mm} perm. \right) S_{\parbox{10mm}{\includegraphics[scale=0.4]{fig21.eps}}} + \mathcal{K}
  \left(\parbox{10mm}{\includegraphics[scale=1.0]{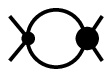}} + 2 \hspace{1mm} perm. \right) S_{\parbox{6mm}{\includegraphics[scale=0.5]{fig25.eps}}}\Biggr],
\end{eqnarray}

\begin{eqnarray}\label{Zphi2}
&& Z_{\phi^{2}}(u,\epsilon^{-1}) = 1 + \frac{1}{2} \mathcal{K} 
\left(\parbox{14mm}{\includegraphics[scale=1.0]{fig14.eps}} \right) S_{\parbox{10mm}{\includegraphics[scale=0.5]{fig14.eps}}} + \frac{1}{4} \mathcal{K} 
\left(\parbox{12mm}{\includegraphics[scale=1.0]{fig16.eps}} \right) S_{\parbox{10mm}{\includegraphics[scale=0.5]{fig16.eps}}}  + \frac{1}{2} \mathcal{K} 
\left(\parbox{11mm}{\includegraphics[scale=.8]{fig17.eps}} \right) S_{\parbox{8mm}{\includegraphics[scale=0.4]{fig17.eps}}} + \nonumber \\ && \frac{1}{2} \mathcal{K}
  \left(\parbox{12mm}{\includegraphics[scale=.2]{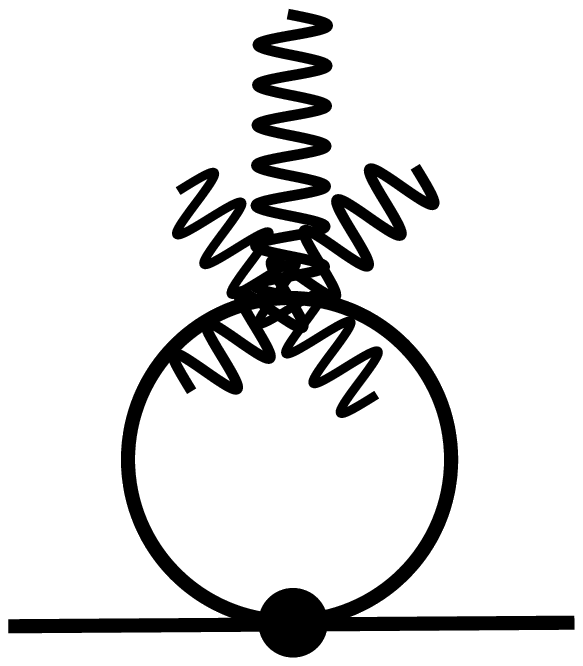}} \right) S_{\parbox{7mm}{\includegraphics[scale=.12]{fig31.eps}}} + \frac{1}{2} \mathcal{K}
  \left(\parbox{12mm}{\includegraphics[scale=.2]{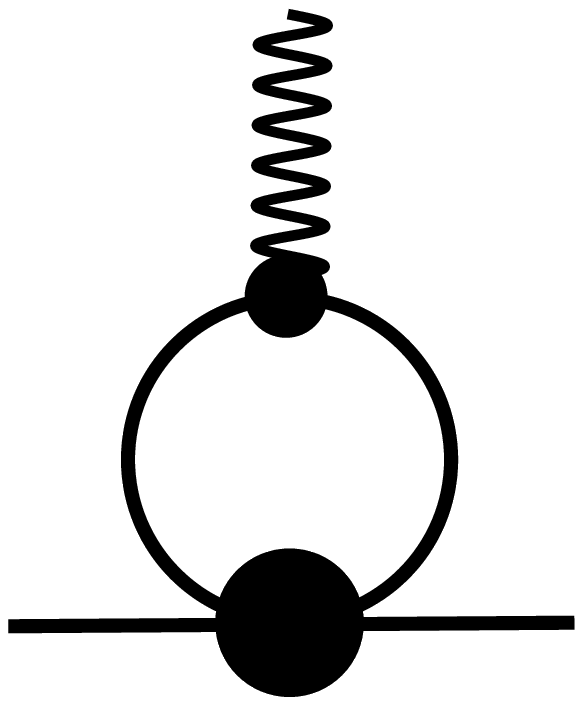}} \right) S_{\parbox{7mm}{\includegraphics[scale=.12]{fig32.eps}}},
\end{eqnarray}
where $S_{\parbox{6mm}{\includegraphics[scale=0.5]{fig6.eps}}}$ is the symmetry factor for the corresponding diagram and so on when some $N$-component field is considered. As it is known, in this method, we do not need to evaluate all diagrams, but just four of them \cite{Int.J.Mod.Phys.B.30.1550259,Amit}. This minimal set of diagrams is composed of the ones
\begin{eqnarray}
&&\parbox{12mm}{\includegraphics[scale=1.0]{fig6.eps}} =  -\frac{u^{2}(P^{2} + K_{\mu\nu}P^{\mu}P^{\nu})}{8\epsilon}\left[ 1 + \frac{1}{4}\epsilon -2\epsilon L_{3}(P^{2} + K_{\mu\nu}P^{\mu}P^{\nu}) \right] \mathbf{\Pi}^{2},
\end{eqnarray}  
\begin{eqnarray}
&&\parbox{10mm}{\includegraphics[scale=0.9]{fig7.eps}} = \frac{u^{3}(P^{2} + K_{\mu\nu}P^{\mu}P^{\nu})}{6\epsilon^{2}}\left[ 1 + \frac{1}{2}\epsilon -3\epsilon L_{3}(P^{2} + K_{\mu\nu}P^{\mu}P^{\nu}) \right] \mathbf{\Pi}^{3},
\end{eqnarray}   
\begin{eqnarray}
&&\parbox{10mm}{\includegraphics[scale=1.0]{fig10.eps}} =  \frac{\mu^{\epsilon}u^{2}}{\epsilon} \left[1 - \frac{1}{2}\epsilon - \frac{1}{2}\epsilon L(P^{2} + K_{\mu\nu}P^{\mu}P^{\nu}) \right] \mathbf{\Pi},
\end{eqnarray}   
\begin{eqnarray}
&&\parbox{12mm}{\includegraphics[scale=0.8]{fig21.eps}} =  -\frac{\mu^{\epsilon}u^{3}}{2\epsilon^{2}} \left[1 - \frac{1}{2}\epsilon - \epsilon L(P^{2} + K_{\mu\nu}P^{\mu}P^{\nu}) \right] \mathbf{\Pi}^{2},
\end{eqnarray}  
where
\begin{eqnarray}\label{uahuahuahlol}
&& L(P^{2} + K_{\mu\nu}P^{\mu}P^{\nu}) =  \int_{0}^{1}dx\ln\left[\frac{x(1-x)(P^{2} + K_{\mu\nu}P^{\mu}P^{\nu})}{\mu^{2}}\right],
\end{eqnarray}
\begin{eqnarray}
&& L_{3}(P^{2} + K_{\mu\nu}P^{\mu}P^{\nu}) =  \int_{0}^{1}dx(1-x)\ln\left[\frac{x(1-x)(P^{2} + K_{\mu\nu}P^{\mu}P^{\nu})}{\mu^{2}}\right]. 
\end{eqnarray}

Now computing the $\beta$-function and anomalous dimensions present in the renormalization group equation
\begin{eqnarray}\left( \mu\frac{\partial}{\partial\mu} + \beta\frac{\partial}{\partial u} - \frac{1}{2}n\gamma_{\phi} + l\gamma_{\phi^{2}} \right)\Gamma^{(n,l)} = 0,
\end{eqnarray}
where 
\begin{eqnarray}\label{kjjaffaxdzs}
\beta(u) = \mu\frac{\partial u}{\partial \mu},\gamma_{\phi}(u) = \mu\frac{\partial\ln Z_{\phi}}{\partial \mu},\gamma_{\phi^{2}}(u) = -\mu\frac{\partial\ln Z_{\phi^{2}}}{\partial \mu},
\end{eqnarray}
we have
\begin{eqnarray}\label{reewriretjgjk}
\beta(u) = -\epsilon u + \frac{N + 8}{6}\mathbf{\Pi} u^{2} - \frac{3N + 14}{12}\mathbf{\Pi}^{2}u^{3},
\end{eqnarray} 
\begin{eqnarray}\label{jkjkpfgjrftj}
\gamma_{\phi}(u) = \frac{N + 2}{72}\mathbf{\Pi}^{2}u^{2} - \frac{(N + 2)(N + 8)}{1728}\mathbf{\Pi}^{3}u^{3},
\end{eqnarray} 
\begin{eqnarray}\label{gfydsguyfsdgufa}
\gamma_{\phi^{2}}(u) = \frac{N + 2}{6}\mathbf{\Pi} u - \frac{5(N + 2)}{72}\mathbf{\Pi}^{2}u^{2}.
\end{eqnarray} 
Computing the nontrivial fixed point and applying the relations $\eta\equiv\gamma_{\phi}(u^{*})$ and $\nu^{-1}\equiv 2 - \overline{\gamma}_{\phi^{2}}(u^{*})$, we obtain that the LV critical exponents are identical to their LV counterparts, thus confirming once again the universality hypothesis. Now we generalize all the results found for finite loop-order for any loop levels.

\section{Exact Lorentz-violating all-loop order critical exponents}

\par For evaluating the critical exponents for all loop levels, we need to apply a theorem \cite{CarvalhoSenaJunior} which permit us to express a general Feynman diagram as $\mathbf{\Pi}^{L}\mathcal{F}(u,P^{2} + K_{\mu\nu}P^{\mu}P^{\nu},\epsilon,\mu)$ if the corresponding LI one has the expression $\mathcal{F}(u,P^{2},\epsilon,\mu,m)$. The number $L$ is the number of loops of referred diagram.  

\par As the critical exponents are the same if computed in any renormalization scheme, we will apply the most general one, the BPHZ method. Now by applying the BPHZ method, for all loop orders, in which all momentum-dependent integrals are eliminated in the renormalization process, order by order in perturbation theory \cite{BogoliubovParasyuk,Hepp,Zimmermann} and the theorem above, a possible LV dependence on $K_{\mu\nu}$ of $\beta$-function and anomalous dimensions coming from the LV momentum-dependent integrals disappears. So, according to the theorem aforementioned, the only LV dependence of these functions on the LV parameter $K_{\mu\nu}$ is the remaining one, that coming from the volume elements of the Feynman integrals, \textit{i. e.}, the LV full $\mathbf{\Pi}^{L}$ factor, where $L$ is the number of loops of the referred term for these functions. Thus we can write the exact LV the $\beta$-function and anomalous dimensions for all loop orders as
\begin{eqnarray}\label{uhgufhduhufdhu}
\beta(u) =  -\epsilon u + \sum_{n=2}^{\infty}\beta_{n}^{(0)}\mathbf{\Pi}^{n-1}u^{n}, 
\end{eqnarray}
\begin{eqnarray}
\gamma(u) = \sum_{n=2}^{\infty}\gamma_{n}^{(0)}\mathbf{\Pi}^{n}u^{n},
\end{eqnarray}
\begin{eqnarray}
\gamma_{\phi^{2}}(u) = \sum_{n=1}^{\infty}\gamma_{\phi^{2}, n}^{(0)}\mathbf{\Pi}^{n}u^{n},
\end{eqnarray}
where $\beta_{n}^{(0)}$, $\gamma_{n}^{(0)}$ and $\gamma_{\phi^{2}, n}^{(0)}$ are the LI nth-loop corrections to the referred functions. Once again, by factoring $u$ from the $\beta$-function of Eq. (\ref{uhgufhduhufdhu}) and evaluating the all-loop nontrivial fixed point, we get the nontrivial solution $u^{*} = u^{*(0)}/\mathbf{\Pi}$, where $ u^{*(0)}$ is the LI nontrivial fixed point valid for all loop levels. Then, we get that the LV critical exponents exponents are identical to their LI counterparts, but now the referred critical exponents are valid for all loop orders.

\section{Conclusions}

\par We computed the all-loop quantum contributions to the critical exponents of LV O($N$) $\lambda\phi^{4}$ scalar field theories, where the Lorentz violation mechanism was treated exactly by keeping the LV $K_{\mu\nu}$ coefficients exactly. Firstly, we analytically and explicitly evaluated, at next-to-leading order, the exact LV radiative quantum corrections to the critical exponents. After that, we generalized these results for all loop levels, through an induction process based on a general theorem emerging from the coordinates redefinition in momentum space directly in Feynman diagrams approach approach. We have obtained that the LV critical exponents were the same as their LI counterparts. The explanation for this fact is that the symmetry breaking mechanism occurs in the space where the field is defined and not in the internal space of the field which, on the other hand, could affect the critical exponents values. Moreover, the LV theory we have considered is only a naively one, as seen through coordinates redefinition techniques \cite{PhysRevD692004105009,PhysRevD832011016013}. Then, the LV theory is converted into an effective Lorentz-invariant one. The present approach, besides exact, can in an evident and straightforward way, furnish correct expressions for the all-loop LV radiative quantum corrections to the $\beta$-function and anomalous dimensions considering just a single concept, namely the loop number of the corresponding term of these functions. We showed that the exact results reduced to its non-exact, earlier obtained, counterparts in the appropriated limit. The present exact approach can inspire the task of considering the exact effect of LV mechanisms in distinct theories in high energy physics field theories of standard model extension for example as well as in low energy physics for condensed matter field theories for considering corrections to scaling and finite-size scaling amplitude ratios as well as critical exponents in geometries subjected to different boundary conditions for systems belonging to the O($N$) and Lifshitz \cite{Phys.Rev.B.67.104415,Phys.Rev.B.72.224432,Carvalho2009178,Carvalho2010151} universality classes etc.

\section{Acknowledgements}

\par With great pleasure the authors thank the kind referee for helpful comments. PRSC and MISJ would like to thank Federal University of Piau\'{i} and FAPEAL (Alagoas State Research Foundation), CNPq (Brazilian Funding Agency) for financial support, respectively.

%\appendix
%
%
%\section{Integral formulas in $d$-dimensional Euclidean momentum space}\label{Integral formulas in $d$-dimensional Euclidean momentum space}
%
%\par 
%
%\begin{eqnarray}
%\int d^{d}q \frac{q^{\mu}}{(q^{2} + 2pq + M^{2})^{\alpha}} = -\hat{S}_{d}\frac{1}{2}\frac{\Gamma(d/2)}{\Gamma(\alpha)}\frac{p^{\mu}\Gamma(\alpha - d/2)}{(M^{2} - p^{2})^{\alpha - d/2}},
%\end{eqnarray}
%
%\begin{eqnarray}
%\int d^{d}q \frac{q^{\mu}q^{\nu}}{(q^{2} + 2pq + M^{2})^{\alpha}} = \hat{S}_{d}\frac{1}{2}\frac{\Gamma(d/2)}{\Gamma(\alpha)}\times \nonumber \\ \left[\frac{1}{2}\delta^{\mu\nu}\frac{\Gamma(\alpha - 1 - d/2)}{(M^{2} - p^{2})^{\alpha - 1 - d/2}} + p^{\mu}p^{\nu}\frac{\Gamma(\alpha - d/2)}{(M^{2} - p^{2})^{\alpha - d/2}} \right].
%\end{eqnarray}

%\section*{Acknowledgements}



%% References


%\bibliography{apstemplate}






% \end{thebibliography}

\end{document}